\documentclass[twocolumn]{aastex63}

\usepackage{amsmath}	
\usepackage{lineno}

\newcommand{\Msun}{M_\odot}
\newcommand{\Mdot}{\dot{M}}
\newcommand{\Mdotacc}{\dot{M}_{\rm acc}}

\newcommand{\Ledd}{L_{\rm Edd}}

\newcommand{\Mdotedd}{\dot{M}_{\rm Edd}}
\newcommand{\ph}{_{\rm ph}}
\newcommand{\kepler}{\texttt{KEPLER}}
\newcommand{\mesa}{\texttt{MESA}}
\newcommand{\yd}{y_{\rm d}}
\newcommand{\yc}{y_{\rm c,min}}
\newcommand{\yw}{y_{\rm w}}
\newcommand{\yej}{y_{\rm ej}}
\newcommand{\h}{^1{\rm H}}

\newcommand{\src}{SAX~J1808.4--3658}
\newcommand{\iso}[2]{^{#1}{\rm #2}}

\shorttitle{Mixed H/He PRE burst lightcurves}
\shortauthors{Guichandut et al.}

\begin{document}
\title{The imprint of convection on Type I X-ray bursts: \\Pauses in photospheric radius expansion lightcurves}

\correspondingauthor{Simon Guichandut}
\email{simon.guichandut@mail.mcgill.ca}

\author{Simon Guichandut}
\affiliation{Department of Physics and Trottier Space Institute, McGill University, 3600 rue University, Montreal, QC, H3A 2T8, Canada}

\author{Andrew Cumming}
\affiliation{Department of Physics and Trottier Space Institute, McGill University, 3600 rue University, Montreal, QC, H3A 2T8, Canada}

\begin{abstract}
Motivated by the recent observation by NICER of a type I X-ray burst from \src{} with a distinct ``pause" feature during its rise, we show that bursts which ignite in a helium layer underneath a hydrogen-rich shell naturally give rise to such pauses, as long as enough energy is produced to eject the outer layers of the envelope by super-Eddington winds. The length of the pause is determined by the extent of the convection generated after ignition, while the rate of change of luminosity following the pause is set by the hydrogen gradient left behind by convection. Using the \mesa{} stellar evolution code, we simulate the accumulation, nuclear burning and convective mixing prior to and throughout the ignition of the burst, followed by the hydrodynamic wind. We show that the results are sensitive to the treatment of convection adopted within the code. In particular, the efficiency of mixing at the H/He interface plays a key role in determining the shape of the lightcurve. The data from \src{} favors strong mixing scenarios. Multidimensional simulations will be needed to properly model the interaction between convection and nuclear burning during these bursts, which will then enable a new way to use X-ray burst lightcurves to study neutron star surfaces.
\end{abstract}


\section{Introduction}\label{sec:intro}
As per the recent MINBAR catalogue \citep{Galloway.Zand.ea2020}, about one fifth of type I X-ray bursts from accreting neutron stars \citep{Lewin.Paradijs.ea1993,Galloway.Keek2021}  reach high enough luminosities to provoke a radiatively-driven expansion of the neutron star envelope. In these ``photospheric radius expansion" (PRE) bursts, the star's photosphere moves outward and appears 10--100 times larger, for a few to tens of seconds. PRE bursts offer a unique opportunity to study not only the surface but also the interior of neutron stars. Indeed, they have been used to place joint constraints on both neutron star mass and radius and the dense matter equation of state \citep[and references therein]{Ozel.Psaltis.ea2016}. Another way to constrain the mass is to measure the gravitational redshift of spectral features from heavy elements being ejected during the burst \citep{Li.Suleimanov.ea2018,Strohmayer.Altamirano.ea2019}. These techniques rely on theoretical models which describe the expansion of the star's envelope and the winds that drive it.

The recent deployment of the Neutron Star Interior Composition Explorer (NICER) telescope has drastically improved observations of PRE bursts, since the instrument's soft X-ray response allows spectral evolution of the PRE to be followed as the blackbody temperature drops to ${\lesssim}1$ keV \citep{Keek.Arzoumanian.ea2018}.
A most interesting PRE burst was recently observed by NICER from the millisecond pulsar SAX J1808.4--3658 \citep{Bult.Jaisawal.ea2019}. During the burst rise, the luminosity briefly ``paused" for ${\approx}0.7$ s before reaching its peak. The ratio between the bolometric luminosity at the peak and pause was ${\approx}1.68$, which is very similar to the ratio between the pure helium and solar composition ($X\approx 0.7$) Eddington luminosities, given by
\begin{equation}
    \Ledd=\frac{4\pi GMm_p c}{\sigma_T(1+X)}\,, \label{eq:Ledd}
\end{equation}
where $M$ is the neutron star mass, $\sigma_T$ is the Thomson scattering cross-section, $m_p$ is the proton mass, and $X$ is the mass fraction of ionized hydrogen (free protons). \citet{Bult.Jaisawal.ea2019} interpreted this as an observation of the rapid ejection of a solar or hydrogen-rich layer, followed by the usual helium PRE phase. This is consistent with the observed burst recurrence times and energetics which indicate that \src\ is in the burst regime where hydrogen is depleted by hot CNO burning well before unstable ignition of helium \citep{Galloway.Cumming2006,Goodwin.Galloway.ea2019}.

A similar idea for a two-staged mixed H/He PRE burst was put forward by \citet{Sugimoto.Ebisuzaki.ea1984} to explain the bursting behaviour of 4U/MXB 1636-53, which showed a bimodal distribution of peak luminosity (see also \citealt{Galloway.Psaltis.ea2006}). Following this suggestion, \citet{Kato1986} computed steady-state solutions of outflows from a small H layer on top of a He-layer, finding the timescale for the ejection of the H layer (and jump in luminosity) to be on the order of 0.1 to 1 s, inversely proportional to the luminosity of the model. However, this strongly depends on not only the mass of the H layer, assumed to be $10^{-16}\Msun$, but also on the steady-state assumption, which cannot reproduce the actual ejection of the H layer. It is clear that in order to understand this type of burst, we need time-dependent hydrodynamic simulations combined with realistic neutron star envelopes.

X-ray bursts are challenging to model because of the many different types of physics involved. Previous studies have followed the time-dependent nuclear burning and convection during the thermonuclear runaway, using stellar evolution codes such as \kepler{} \citep{Woosley.Heger.ea2004,Cyburt.Amthor.ea2010}, or Modules for Experiments in Stellar Astrophysics, \mesa{}, \citep{Paxton.Bildsten.ea2011,Meisel2018}, but did not resolve the formation of the wind in PRE bursts.
\citet{Yu.Weinberg2018} first demonstrated the ability of \mesa{} to resolve both the nuclear burning and convective mixing at the onset of bursts from pure He accretion, followed by the hydrodynamic ejection of a super-Eddington wind and the PRE phase. This opened up the possibility of simulating time-dependent PRE bursts with an emphasis on the role of composition in the resulting wind.

In this paper, we use \mesa{} \citep[version 15140;][]{Paxton.Bildsten.ea2011,Paxton.Cantiello.ea2013,Paxton.Marchant.ea2015,Paxton.Schwab.ea2018,Paxton.Smolec.ea2019} to simulate the accumulation phase, ignition, super-Eddington wind, and decay of a mixed H/He PRE burst. This represents the first full simulation of PRE bursts resulting from accretion of H and He. Our main result is that the rise in luminosity at the start of the burst pauses temporarily once the luminosity reaches the Eddington luminosity. As a wind develops and mass is ejected, deeper layers are eventually exposed that have been depleted in hydrogen by a combination of convective mixing and nuclear burning. This ends the pause and the luminosity begins to rise again as the outflowing material becomes less hydrogen rich and therefore has a larger Eddington luminosity. The resulting lightcurve has a distinct shape in contrast to pure helium bursts, and it depends on the gradient of hydrogen left behind by convection. However, as we will show, these results are very sensitive to the treatment of convection within the code. Indeed, the entrainment of protons by convection from the H layer into the He burning zone below leads to short timescale nuclear burning, which feeds back into the convection. This is a regime that cannot be adequately simulated in one dimension, making the detailed shape of light curve uncertain.

We begin in Section~\ref{sec:description} with a simple model of mixed H/He PRE bursts, and explain the connection between the hydrogen profile in the envelope post-convection and the shape of the lightcurve. The main stages and important parameters of this model are also summarized in Figure~\ref{fig:diagram}. In Section~\ref{sec:mesa}, we describe our \mesa{} simulations and show detailed results using the simplest prescription for convection. In Section~\ref{sec:otherconvection}, we vary the prescription for convection and assess how the results are affected. In Section~\ref{sec:discussion}, we summarize our findings, elaborate on issues related to the treatment of convection in one-dimensional simulations, and give an interpretation for the \src{} data.

\section{Evolution of the composition profile and lightcurve shape}\label{sec:description}
\begin{figure*}[ht]
    \centering
    \includegraphics[width=\linewidth]{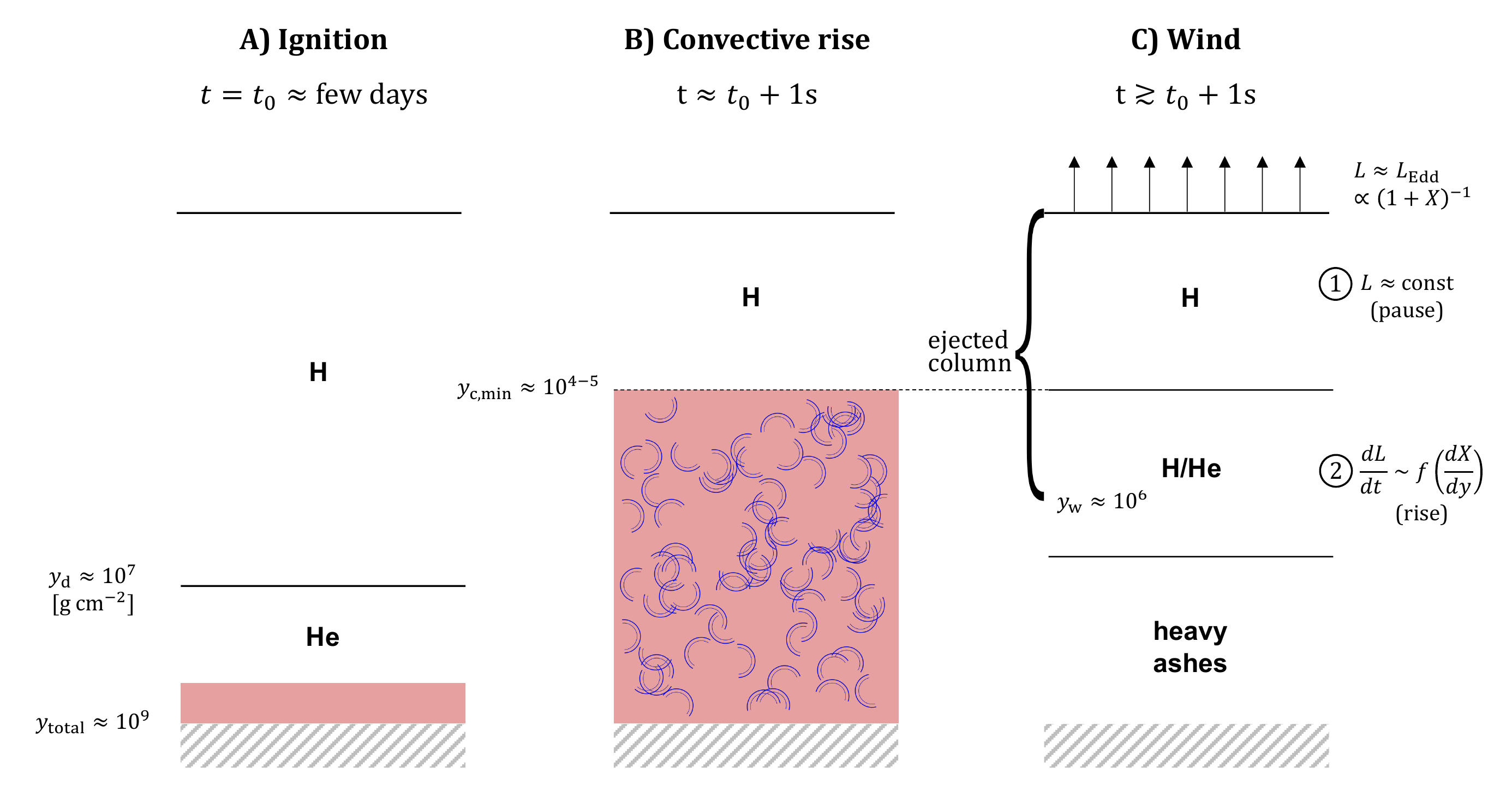}
    \caption{Diagram illustrating the three main stages of the mixed H/He burst. \textit{A)} As hydrogen burns stably throughout accretion, distinct hydrogen and helium-rich layers build up in the envelope, their boundary being at a known column depth of $\yd$ (Equation~\ref{eq:yd}). We are interested in bursts that ignite at $y>\yd$ (red color indicates nuclear burning). \textit{B)} Heat from nuclear burning creates a growing convection zone (blue semicircles) which penetrates into the H-rich layer, resulting in additional nuclear burning. \textit{C)} As the convection zone retreats, it leaves behind a layer of constant hydrogen fraction at a column $\yc<\yd$, a mixed H-He layer and nuclear ashes at depth. Winds progressively eject the layers, up to a column $\yw>\yc$, during which the observed luminosity at infinity is the Eddington luminosity, which depends on the hydrogen fraction $X$ of the material. Since $X$ is initially constant, we first observe a pause after the initial burst rise. After the H layer is ejected, the luminosity once again rises in a manner that depends on the hydrogen gradient $dX/dy$. See text for further details.}
    \label{fig:diagram}
\end{figure*}

The structure of the neutron star envelope at ignition is determined mainly by the accretion rate onto the neutron star and composition of the infalling gas. For $\Mdotacc\gtrsim 2\times 10^{-10} \Msun$ yr$^{-1}$, hydrogen burns via the hot CNO cycle at a constant rate \citep{Bildsten1998}. Then, it can be shown that the column depth, $y(r)\equiv\int_\infty^r \rho(r') dr'$, at which hydrogen is depleted is
\begin{equation}\label{eq:yd}
    \yd=2.7\times 10^7\,\rm{g}\,{\rm cm}^{-2}\left(\frac{\Mdotacc}{0.01\Mdotedd}\right)\left(\frac{0.02}{Z_{\rm CNO}}\right)\left(\frac{X_0}{0.7}\right)\,
\end{equation}
where $X_0$ and $Z_{\rm CNO}$ are the initial hydrogen and CNO nuclei mass fractions \citep{Cumming.Bildsten2000}\footnote{The expression for $\yd$ from \citet{Cumming.Bildsten2000} was re-evaluated for a 12 km radius and using a more accurate $E_{\rm H}\approx 6.0\times 10^{18}$ erg g$^{-1}$, instead of 6.4, to account for neutrino losses \citep{Wallace.Woosley1981}.}. We scale the accretion rate to the Eddington accretion rate corresponding to a neutron star with $R=12$ km and accreted hydrogen mass fraction $X_0=0.7$, giving $\Mdotedd=8\pi Rm_pc/(1+X_0)\sigma_T=2.1\times10^{-8} \Msun$ yr$^{-1}$. Therefore at ignition, the envelope consists of two layers: an outer H-rich layer of depth $\yd$ in which the hydrogen abundance drops from the accreted value to zero, and an inner layer of pure helium where ignition of the burst occurs. This initial state is illustrated in column A of Figure~\ref{fig:diagram}.

From an energetics standpoint, we know that only a column $\yw\sim10^6\text{--}10^7$ g cm$^{-2}$ can be ejected by winds \citep{Weinberg.Bildsten.ea2006}, which is smaller than $\yd$. However, \citet{Weinberg.Bildsten.ea2006} also showed that prior to the wind, a convection zone will grow and extend to a column smaller than $\yw$ (Figure~\ref{fig:diagram} column B), which will mix the H and He and result in a change in the composition of the ejecta as a function of time. We define the minimum column depth reached by convection as $\yc$, below which hydrogen is not mixed and $X=X_0$ is roughly constant\footnote{$X$ in fact decreases linearly with $y$, but since $\yc$ ends up being ${\sim}1\%$ of $\yd$ or less, the variation in $X$ over this column is negligible.}. 

Wind models for PRE bursts \citep{Ebisuzaki.Hanawa.ea1983,Paczynski.Proszynski1986,Joss.Melia1987,Guichandut.Cumming.ea2021} show that the luminosity at infinity  is always very close to the Eddington luminosity $\Ledd$ (Equation~\ref{eq:Ledd}), as any ``extra" energy in the form of a super-Eddington flux gets used up to drive mass-loss. Therefore, during the initial ejection of $\yc$, $L\approx \Ledd$ will be constant. This is the observed pause, and its duration is
\begin{align}
    \Delta t_{\rm p}&\approx \frac{4\pi R^2 \yc}{\Mdot}\nonumber\\
    &\approx 0.18\,\rm{s}\,\left(\frac{\yc}{10^4 \,\rm{g\,cm}^{-2}}\right)\dot{M}_{18}^{-1}\,,\label{eq:tpause}
\end{align}
where $\dot{M}_{18}=\Mdot/(10^{18}$ g s$^{-1}$) is the mass-loss rate, which we assume to be constant during the pause.

After the pause, the ejection of the mixed layers will begin, and the luminosity will rise as the hydrogen fraction $X$ in the ejecta decreases. The rate $dL/dt$ at which the luminosity will increase, assuming it stays near Eddington, will be proportional to $dX/dt\propto \Mdot (dX/dy)$, thus linking the shape of the lightcurve to the hydrogen gradient in the envelope. Column C of Figure~\ref{fig:diagram} illustrates the compositional nature of these two stages, the pause and the rise. Note that if $X=0$ at columns $y<\yw$, a third stage will appear where the luminosity peaks at the helium $\Ledd$ and remains there for the rest of the PRE (until all of $\yw$ has been ejected).

The burst lightcurve is therefore determined by \textit{1)} the nuclear burning and convection that occur during the rising phase, setting the hydrogen profile, and \textit{2)} the mass-loss rate during the wind phase. In the next section,  we describe simulations with \mesa{} to investigate both of these factors.

\section{\mesa{} Simulations}\label{sec:mesa}
  \begin{figure*}[ht!]
    \centering
    \includegraphics{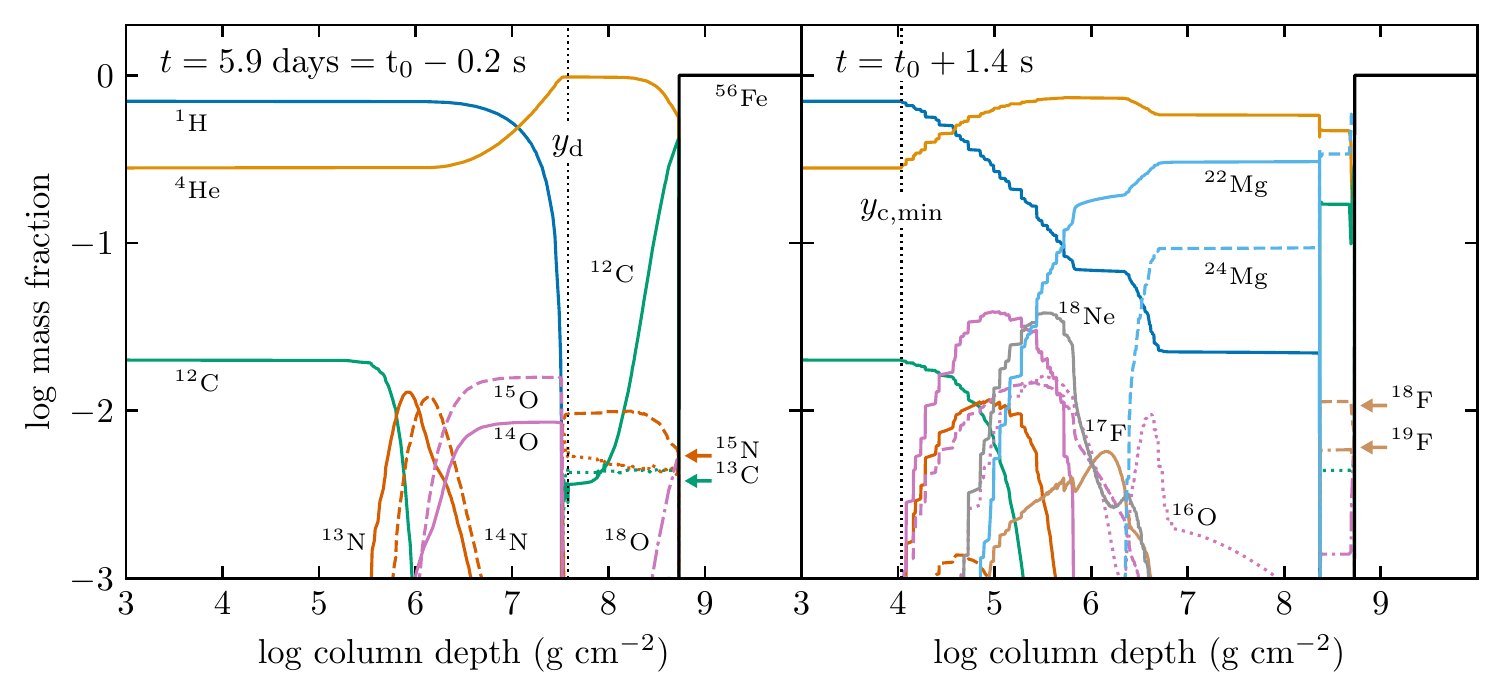}
    \caption{Composition profiles before (left) and after (right) the thermonuclear flash, which ignites at $t=t_0$. The lines for different elements have the same color in left and right panels. In the ${\sim}1.4$~s to reach the Eddington luminosity, convection has significantly mixed the He and H layers. Dotted lines show the depletion column $\yd$ (left) and the minimum extent of convection $\yc$ (right). All runs shown in this paper begin at the ignited model $t=t_0$, which the left panel leads up to. The panel on the right shows the result of mixing using the Schwarzschild criterion prescription for convective boundaries (see Section~\ref{subsec:ignition}).}
    \label{fig:composition}
\end{figure*}

We model a single burst with several separate \mesa{} runs\footnote{\label{footnote:zenodo}Our \mesa{} inlists, models and simulation setup are available at \dataset[10.5281/zenodo.8048553]{https://doi.org/10.5281/zenodo.8048553}.}, in an approach similar to \citet{Yu.Weinberg2018}. First, we follow the ignition and convective rise of the burst under the assumption of hydrostatic equilibrium, then use \mesa{}'s hydrodynamic solver to follow the super-Eddington wind phase. However, unlike \citet{Yu.Weinberg2018}, we leave nuclear burning on during the wind, as much of the energy in bursts with hydrogen comes from slower reactions that continue into the wind phase.

 \subsection{Accumulation and formation of the layers}

The basic physical setup is the same for all simulations: we assume a non-rotating neutron star of mass $M=1.4\Msun$ and radius $R=12$ km, and ignore general relativistic corrections. The envelope initially consists of an $^{56}$Fe substrate\footnote{To build the starting model, we took the the \texttt{ns\_env} model file from the \texttt{ns\_he} problem provided as part of \mesa{}'s test suite, then relaxed the neutron star radius from 10 to 12 km, and finally accreted additional $\iso{56}{Fe}$ to the target column depth. The point of increasing the mass of the iron substrate is to build a large enough buffer between the flashing zone and the inner boundary, allowing heat to diffuse inward without reflecting back.} with a column depth $y=10^{11}$ g cm$^{-2}$. The outer boundary is initially set at an optical depth $\tau=100$ to avoid numerical issues caused by radiation-dominated regions becoming convective \citep{Paxton.Cantiello.ea2013}\footnote{In this first part of the simulation, the outer layers do not expand significantly and we mainly focus on the effects of convection at large depths. We later relax this outer boundary to a more appropriate value to model the wind and lightcurve (Section \ref{subsec:wind}).}. We begin accreting a solar-like composition ($\iso{1}{H}$, $\iso{4}{He}$ and $\iso{12}{C}$ with mass fractions $X=0.7$, $Y=0.28$ and $Z=Z_{\rm CNO}=0.02$ respectively) at a constant rate of $\Mdotacc=3\times10^{-10}\;\Msun$ yr$^{-1}=0.014\Mdotedd$. We assume carbon to be the only metal being accreted for simplicity. What matters for the ignition of this type of burst is to achieve hydrogen depletion, and any isotope part of the CNO cycle would work, because the CNO abundances quickly 
adjust to the equilibrium ratio of $\iso{14}{O}$ and $\iso{15}{O}$ in the hot CNO cycle \citep{Bildsten1998}\footnote{We assume that all of the metallicity is in CNO elements. A more realistic solar composition would also contain non-CNO species, which would reduce the hot CNO burning rate. However, the change in the hydrogen depletion and ignition depths would be smaller than that associated with the uncertain accretion rate and crust heating parameter.}. Throughout accretion, the luminosity at the base of the substrate is fixed to $1.8\times10^{34}$ erg s$^{-1}$, equivalent to $\Mdotacc$ times 1 MeV per nucleon which is roughly the expected crust heating at low accretion rates \citep{Brown2000}. As discussed in \citet{Yu.Weinberg2018}, $\Mdotacc$ and $L_{\rm base}$ determine the ignition depth, which we have chosen to be at $y\approx 3\times 10^8$ g cm$^{-2}$, greater than $y_d$ so that the burst ignites in a pure He layer.

To reduce the complexity of the computations, especially during the wind phase, we use {\mesa}'s \texttt{cno\_extras.net} nuclear network to model the nuclear burning, which includes 17 isotopes up to $^{24}$Mg (we also add $^{56}$Fe as an additional inert element). This is fewer and lighter isotopes than the \texttt{approx21.net} network used by \citet{Yu.Weinberg2018} for He bursts (which has $\alpha$-capture reactions up to iron-group elements), but contains a similar number of reactions due to the addition of hot CNO. It is also a much smaller network than that used by \citet{Woosley.Heger.ea2004}, who studied the energy generation carefully but did not model the hydrodynamic wind. Since we are focused primarily on the wind, this limited network is adequate because most of the energy generated during the burst comes from hydrogen and helium burning in CNO and triple-$\alpha$ reactions. However, this assumption means that our calculations do not accurately predict the nuclear burning ashes, both ejected in the wind and leftover afterwards.

The left panel of Figure~\ref{fig:composition} shows the composition profile of the envelope after 5.9 days of accretion. The infalling CNO species convert to equilbrium oxygen ratios in under an hour. The hydrogen depletion column is $\yd=3.7\times10^7$ g cm$^{-2}$, consistent with Equation~\ref{eq:yd} for the chosen $\Mdotacc$. At larger column depths, the CNO cycle is starved of protons, and the abundances are determined by $\beta$-decay rates only. Prior to ignition, some helium has already started stably burning and converting to $^{12}$C.

\begin{figure*}[ht]
    \centering
    \includegraphics{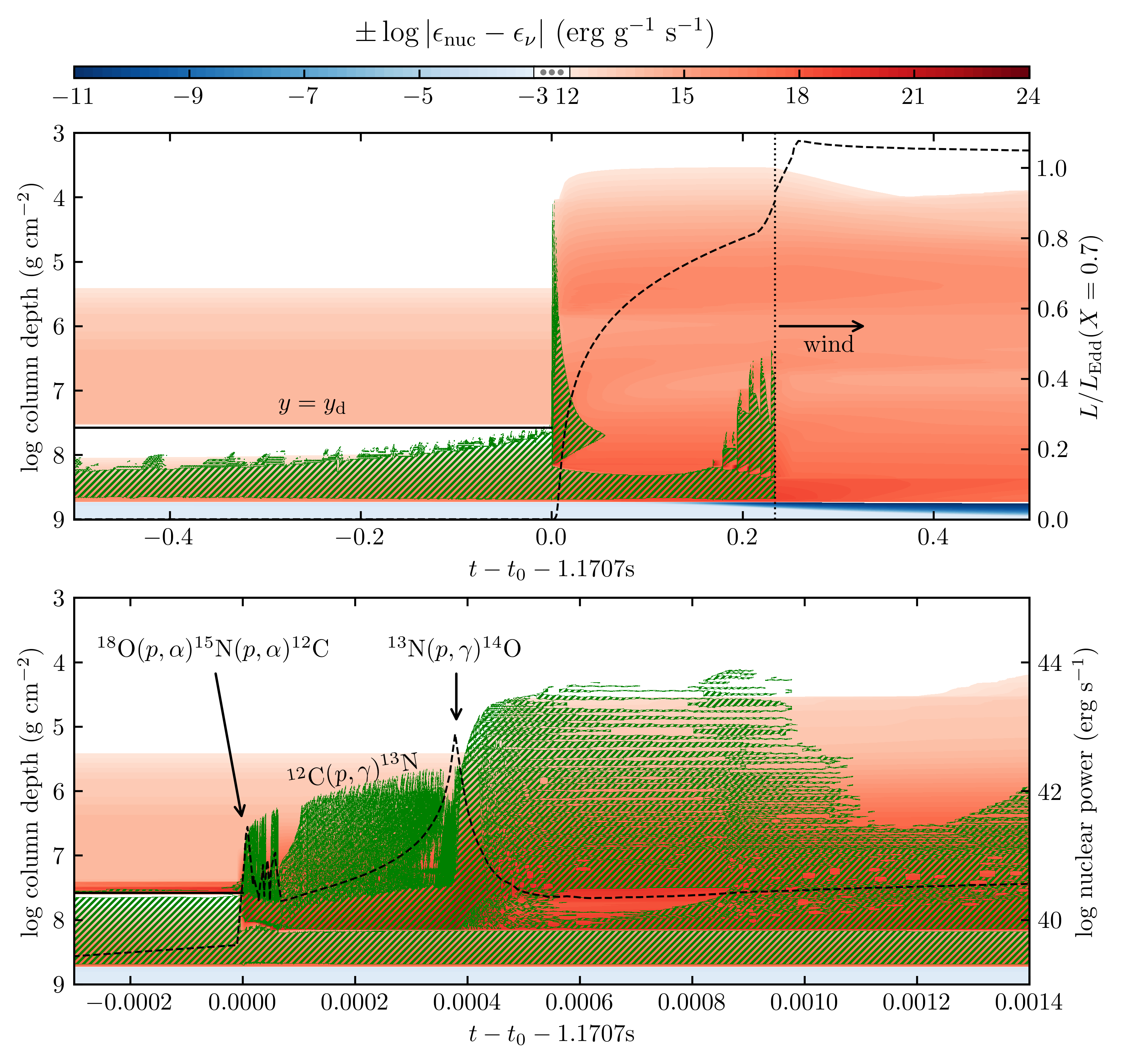}
    \caption{Kippenhahn diagrams for the Schwarzschild run, centered on the moment of collision between the convection zone and the H layer. The bottom panel is zoomed into a 1.5~ms window following the collision. The color scale traces the energy generation or loss (nuclear burning minus all neutrino losses), while the green hatches mark the convection zones. The solid black lines show the location of the depletion depth $\yd$. In the top panel, the dashed line shows the luminosity coming out of the atmosphere, normalized by the Eddington luminosity of the accreted gas (scale on the right-hand side). In the bottom panel, it shows the integrated nuclear power ($\int dm\, \epsilon_{\rm nuc}$). Moments where certain reactions dominate are labeled (see Section~\ref{subsec:ignition}).}
    \label{fig:kipp}
\end{figure*}

\subsection{Ignition and convective rise}\label{subsec:ignition}

\begin{figure*}[ht]
    \centering
    \includegraphics{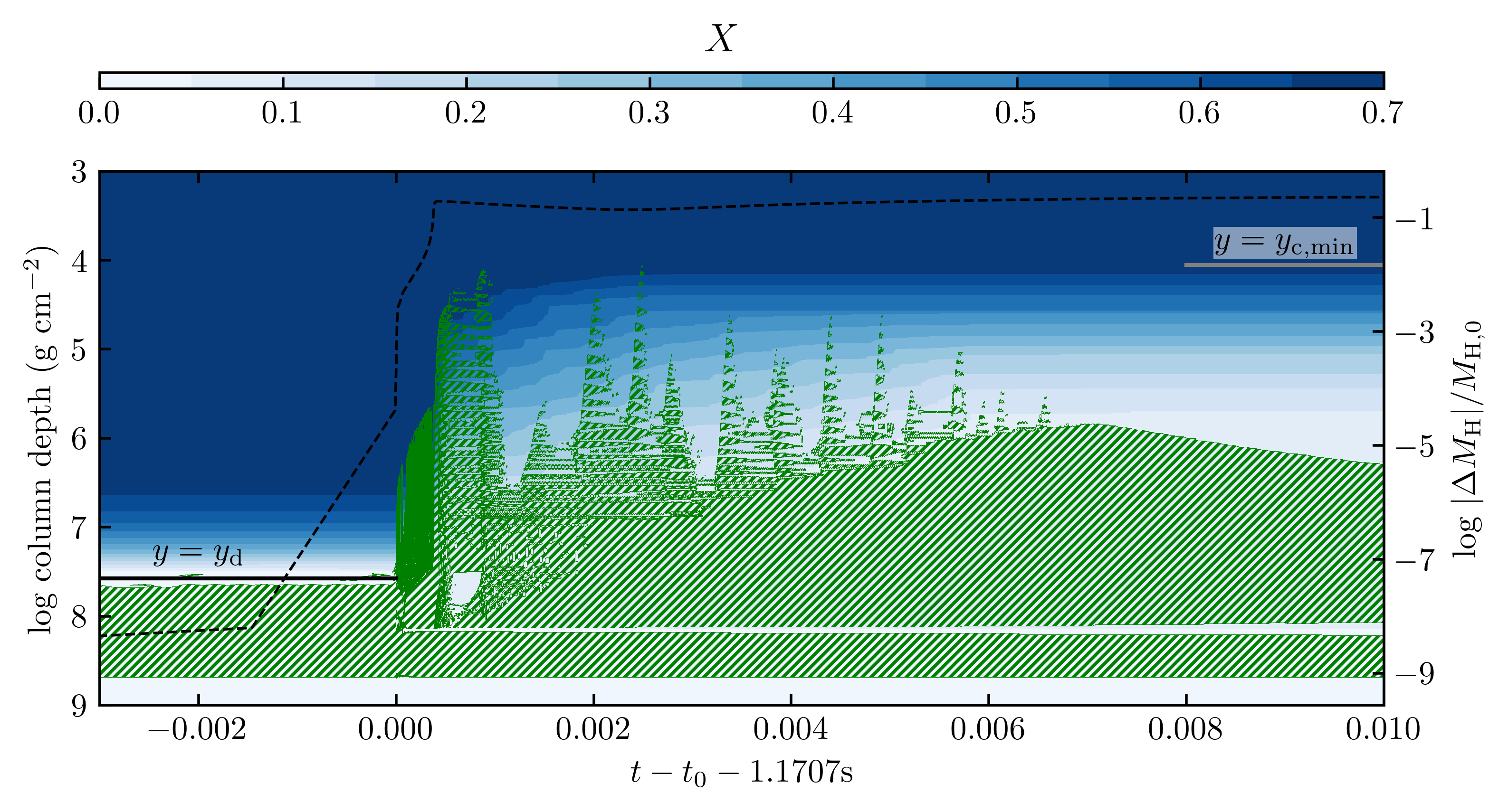}
    \caption{Kippenhahn diagram for the Schwarzschild run as in Figure~\ref{fig:kipp}, now with the color scale representing the hydrogen abundance. The solid black and gray lines show $\yd$ and $\yc$ (same values as in Figure~\ref{fig:composition}). The dashed line shows the relative change in the total mass of hydrogen at a given time compared to the initial amount at ignition (scale on the right-hand side).}
    \label{fig:kippH}
\end{figure*}

After having built up the He layer, unstable triple-$\alpha$ burning triggers the thermonuclear runaway. We define ‘‘ignition'' as the moment $t=t_0$ when the He layer is convective and has a larger maximum nuclear energy generation rate than the H layer. The convection zone begins growing in the He layer as it would in a pure He burst, then hits the H layer about 1.1s later. The mixing of H into the convection zone leads to a sudden increase in the nuclear energy generation rate at the top of the zone. The outcome of this mixing event depends strongly on the prescription used in the code for convection. Here, we use the prescription of \citet{Henyey.Vardya.ea1965} for mixing length theory, with the dimensionless parameter $\alpha_{\rm mlt}=1.5$ dictating the ratio between the mixing length and local pressure scale height. In this section, we present results using the Schwarzschild criterion to determine convective boundaries (as assumed also by \citealt{Yu.Weinberg2018}). This ignores the effects of composition gradients on the convective stability, but simplifies the interpretation of our results. We explore the effect of changing the prescription for convection in Section~\ref{sec:otherconvection}.

In Figure~\ref{fig:kipp}, we show Kippenhahn diagrams for the history of convection and nuclear burning as a function of depth, throughout the flash phase and beginning of the wind. The bottom panel is zoomed in to show short timescales following the collision between the He convection zone and the H layer. In this collision, fresh protons are brought in below the depletion depth $\yd$ where they can capture onto seed nuclei, causing a rapid injection of energy and a local steepening of the temperature gradient, which in turns drives further expansion of the convection zone. The first proton captures to happen are $\iso{18}{O}(\rm{p},\alpha)\iso{15}{N}$ and $\iso{15}{N}(\rm{p},\alpha)\iso{12}{C}$. The remaining protons (and fresh ones coming from the top) then quickly capture onto the carbon and build up $\iso{13}{N}$. The nuclear reactions in this first stage do not produce enough heat to generate large scale convection. Instead, the convection splits into many zones, as radiative gaps as small as 0.1\% of the scale height appear. These zones and gaps are clearly seen in the bottom panel of Figure~\ref{fig:kipp}. The maximum number of individual convective zones is 66, and it occurs 0.1 ms after the collision. The maximal extent of the convection during this initial stage is to a column ${\approx}10^6$ g cm$^{-2}$. We later refer to this stage as the \textit{precursor}.

About 0.4 ms after the initial collision, enough nitrogen has built up to trigger a ``second ignition" via the $\iso{13}{N}(\rm{p},\gamma)\iso{14}{O}$ reaction. This time, so much heat is released that the convection zone grows massively in less in 0.1 ms, its column depth extent decreasing by a factor of ${\approx}30$. The convection is still split but the radiative gaps now maintain over time, resulting in a period of layered convection, which extends down to a minimum column $\yc= 1.1\times10^4$ g cm$^{-2}$. The remnants of this period of rapid burning appear in the final composition profiles (see e.g. the magenta and orange lines for $\iso{14}{O}$ and $\iso{13}{N}$ respectively in the right panel of Figure~\ref{fig:composition}), and will later be partly ejected by the wind. Further oxygen, neon and finally sodium-burning reaches the end-point of our network at $\iso{24}{Mg}$. The final composition profiles that go into the hydrodynamic calculation are shown in the right panel of Figure~\ref{fig:composition}. If a more complete network were to be used, we would expect the ashes to proton and $\alpha$ capture to heavier elements, namely Ca and Si, over the following $\sim$tens of seconds \citep{Woosley.Heger.ea2004}.

At the onset of both of these stages, the precursor and $\iso{13}{N}$ ignition, convective velocities on the order of $10^{7}$ cm s$^{-1}$ are generated, which allow for the overall convective envelope (disregarding gaps) to expand by meters in tens of microseconds. In most of the envelope, the large temperatures are such that these velocities remain well below the sound speed. However, in some convective zones, the local Mach number does reach up to ${\sim}15\%$. This implies that the rapid burning is approaching a dynamical regime, but not so much as to make the hydrostatic assumption invalid. We return to the question of convective velocities in Section \ref{sec:otherconvection}.

As we have explained in Section~\ref{sec:description}, the main predictor for the shape of the lightcurve is the hydrogen gradient left over by convection. We now investigate what creates this gradient. Figure~\ref{fig:kippH} is another Kippenhahn diagram of the same simulation that shows the evolution of the hydrogen abundance after the collision. A few ms after the collision, once the convection retreats, the hydrogen gradient is already set. The dashed line traces how much hydrogen has burned away since ignition. At the collision, a significant amount of protons capture onto seed nuclei, and this continues throughout the burst even as the convection zone retreats. At ignition, the total mass of the envelope (minus the iron substrate) is ${\sim}10^{22}$ g, and ${\sim}2$\% of it is H. By the time the gradient is set, about 25\% of the hydrogen has burned away (another 10\% burns in the rest of the flash, mostly at depths $y>\yw$, before ejection by winds). What is important in determining the gradient is not the details of the nuclear burning; once convection subsides, hydrogen burning at shallow depths (near $\yc$) is slow and does not substantially affect the gradient and therefore the lightcurve. Instead, it is the efficiency of the convective mixing which determines how fast and how far protons can be brought downwards to regions of high temperatures where they can quickly burn away. We further investigate the point about the efficiency of mixing in Section~\ref{sec:otherconvection}.

\subsection{Wind and collapse}\label{subsec:wind}

As the outgoing luminosity rises and approaches the Eddington limit a short time after ignition, the outer layers become radiation pressure dominated and the envelope begins to expand. In Newtonian gravity, we know from previous work that appreciable expansion of ${\sim}100$~m above the stellar surface occurs at ${\approx}90\%$ of $\Ledd$ \citep[see Figure 12 of][]{Guichandut.Cumming.ea2021}. At this point, we turn off convection\footnote{This is done to avoid complications associated with radiation-dominated artificially becoming convective.} and accretion, turn on \mesa{}'s hydrodynamics calculation, and relax the outer boundary to an optical depth $\tau=2/3$ in order to resolve the photosphere, as in \citet{Yu.Weinberg2018}. Mass-loss is then done by removing any grid points with a density $\rho<\rho_{\rm thresh}=10^{-7}$ g cm$^{-3}$. To avoid issues caused by going off the opacity tables at low density, we switch to an interpolation formula for electron scattering opacity as a function of temperature from \citet{Paczynski1983}. This is a good approximation at the high temperatures of the wind where electron scattering dominates. 

The presence of jumps in the $\iso{1}{H}$ mass fractions  as a result of convection (see right panel of Figure~\ref{fig:composition}) added some numerical difficulties in the simulation of the winds. Since the acceleration of a fluid parcel due to radiation is proportional to its opacity, a jump in the hydrogen fraction X will result in a density inversion, as the uppermost fluid element is ejected faster than the bottom one can follow. We found that these density inversions tended to expand as they moved outwards, and caused the \mesa{} integration to diverge as they approached $\rho_{\rm thresh}$. A solution that worked in all cases was to manually soften those composition jumps using a smoothing spline on the $^1$H mass fractions prior to running the wind simulation, as shown in the top panel of Figure~\ref{fig:h1_lc}. In order to preserve the overall gradient, this smoothing was done using monotonic functions. The $^4$He profiles were then adjusted such that the sum of mass fractions of all species remained 1 everywhere. 

\begin{figure}
    \centering
    \includegraphics{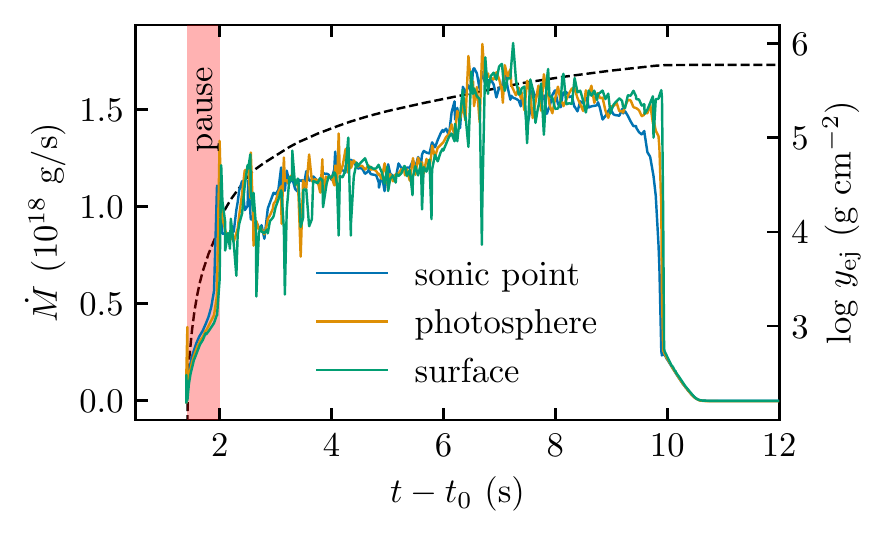}
    \caption{The mass-loss rate of the wind as a function of time for the Schwarzschild run, in solid lines. The different colors show $\Mdot$ evaluated with Eq.\eqref{eq:Mdot} at different locations (see text). The dashed lines show the ejected column $\yej$ (scale on the right-hand side). The red shaded region marks the pause in the lightcurve (bottom panel of Figure~\ref{fig:h1_lc}).}
    \label{fig:Mdot}
\end{figure}

\begin{figure}
    \centering
    \includegraphics{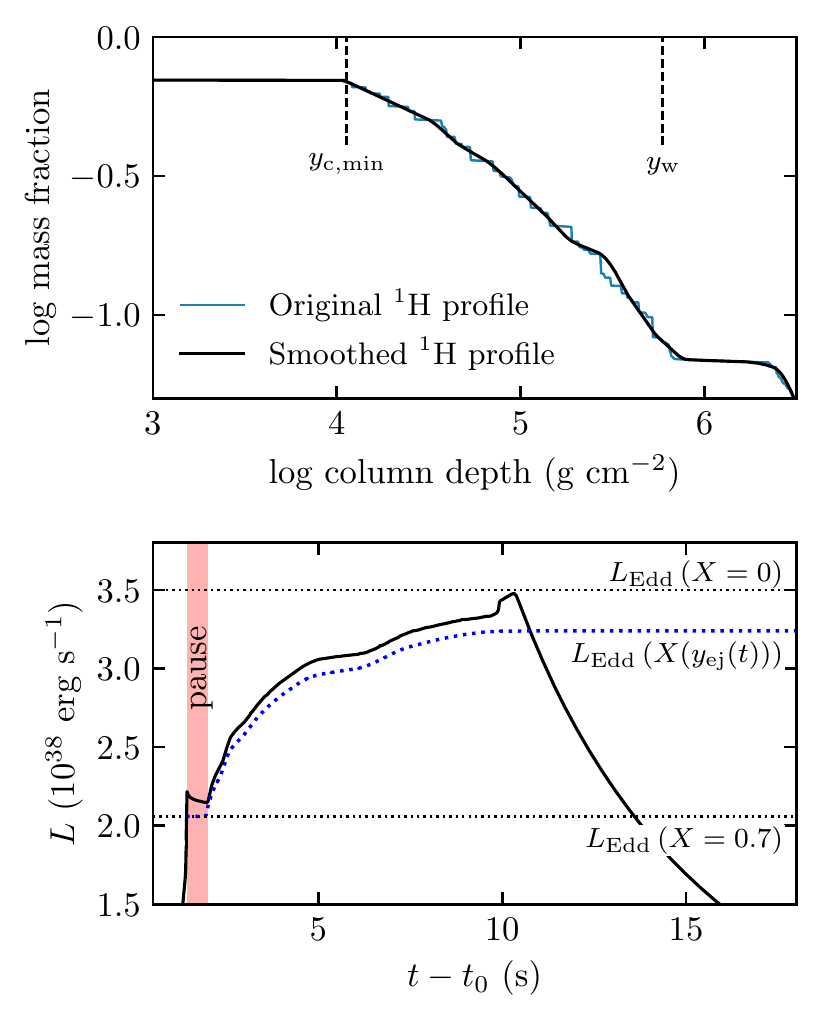}
    \caption{Results of the Schwarzschild run. \textit{Top:} Hydrogen profile after convective mixing. To avoid numerical difficulties during the hydrodynamic wind (see Section~\ref{subsec:wind}), the profile was smoothed using a monotonic cubic spline. The dashed lines show the minimum column reached by convection $\yc$, and the total column ejected by winds $\yw$. \textit{Bottom:} Lightcurve of the burst. The pause occurs after surpassing $\Ledd$ of the accreted material ($X=0.7$, bottom dotted line). The following rise takes place over a much longer timescale since $\yw{\gg}\,\yc$. In general, the outgoing luminosity follows the Eddington luminosity of the material which is currently being ejected (blue dotted line), which we can track using $\yej(t)$ (see Section~\ref{subsec:lightcurves}).}
    \label{fig:h1_lc}
\end{figure}

During the wind, the rate of change of composition in the atmosphere is determined by the mass-loss rate
\begin{equation}\label{eq:Mdot}
    \Mdot(r,t)=4\pi r^2\rho(r,t)v(r,t)\,,
\end{equation}
where $\rho$ and $v$ are the gas density and velocity at a radial distance $r$ from the center of the star. We evaluated $\Mdot$ at three different locations: a) the ``sonic point", i.e.~where the velocity $v=\sqrt{kT/\mu m_p}$ where $\mu$ is the mean molecular weight of the gas, b) the photosphere, i.e.~the location $r=r\ph$ where the luminosity $L=4\pi r\ph^2\sigma T^4$, and c) the surface of the model or outer boundary of the grid. As shown in Figure~\ref{fig:Mdot}, despite some small variations, the mass-loss rate is overall constant across all locations. This is no surprise, as we expect these winds to reach a steady-state, characterized by a constant $\Mdot(r)$ at fixed $t$, in a time much shorter than the evolution timescale of the burst \citep{Joss.Melia1987,Guichandut.Cumming.ea2021}. We can then write the total column ejected as a function of time,
\begin{equation}\label{eq:yej}
    y_{\rm ej}(t)=\frac{1}{4\pi R^2}\int_0^t\Mdot(t')dt'\,,
\end{equation}
independently of location. The total ejected column $\yw$ is the final value of $\yej$, equal to $5.83\times 10^{5}$ g cm$^{-2}$ in this simulation (see dashed line in Figure~\ref{fig:Mdot}). This is roughly consistent with results obtained by \citet{Yu.Weinberg2018} for the total ejected mass of pure He bursts igniting at a similar column depth as we have here. As shown by these authors, the total ejected column, and therefore the total duration of the PRE, would increase (decrease) for bursts which ignite at larger (smaller) column depths.

Approaching the end of the super-Eddington phase of the burst, the nuclear luminosity which is driving the wind begins to die down. The outflow then separates into two regions, an outer unbound wind of high velocity which is being ejected, and an inner atmosphere which is collapsing back into, eventually, a hydrostatic configuration. This can be seen from Figure~\ref{fig:Mdot} at $t-t_0>9$~s, where the mass-loss rate first drops sequentially from the inside (lower radii first). The timescale for the collapse to reach the surface from the sonic point is ${\lesssim}1$~s, which is roughly the sound crossing time between those two locations \citep{Guichandut.Cumming.ea2021}. In some other simulations (Section~\ref{sec:otherconvection}), the collapse results in numerical issues as the infalling gas becomes supersonic, which causes the time-step to drop and the evolution of the model to come to a stop. The moment where these numerical issues arise coincides with the sonic point crossing the photosphere and the wind effectively becoming optically thin. Then, the implicit optically thick assumption made by \mesa{} to treat the radiative transfer becomes incorrect. Future work is needed to investigate the infall phase in more detail.

\subsection{Lightcurve}\label{subsec:lightcurves}

In order to plot the observed lightcurve, we evaluate the radiative luminosity of our models as a function of time at the photosphere. We show the lightcurve for the main Schwarzschild run in the bottom panel of Figure~\ref{fig:h1_lc}. Its shape is consistent with the hydrogen profile in the envelope post-convection, shown in the top panel. On the luminosity axis, the ratio between the peak and pause luminosities is ${\approx}1.6$, consistent with the ratio of Eddington luminosities $(1+X_0)/(1+X(\yw))$ with $X(\yw)=0.07$ (Figure~\ref{fig:h1_lc} top panel). On the time axis, we also have to take into account how the mass-loss rate varies throughout the burst. The pause duration is 0.6~s which, for $\yc=1.1\times 10^4$~g cm$^{-2}$, corresponds to a mass-loss rate $\Mdot_{18}=0.33$ according to Equation~\ref{eq:tpause}. This is in good agreement with the time-averaged value of $\Mdot$ during the pause, $3.5\times 10^{17}$ g s$^{-1}$ (note however that $\Mdot$ changes significantly during the pause, from ${\approx}6.5\times10^{16}$ g s$^{-1}$ to ${\approx}7.6\times 10^{17}$ g s$^{-1}$, see Figure \ref{fig:Mdot}). If the mass-loss rate remained at this value throughout the remainder of the wind, the total duration of the PRE phase would be $(\yw/\yc)\Delta t_{\rm pause}\approx 32$~s. But since $\Mdot$ increases by a factor of ${\sim}3$ after the pause (Figure~\ref{fig:Mdot}), the ejection is much faster and the PRE only lasts ${\approx}9$ s.

As expected, the lightcurve can be obtained by tracking $\Ledd$ as the hydrogen mass fraction in the ejecta $X(\yej(t))$ evolves in time. This is shown by the blue dotted line in Figure~\ref{fig:h1_lc} which closely follows the luminosity from the simulation (black line). Comparing the two, we see that two features of the observed lightcurve are unexplained by composition changes. First, the luminosity during the pause is not exactly flat, but instead slowly decreases throughout its duration. This effect can be understood by considering the energetics of the expansion. At the beginning of the pause, the wind is not yet established. To do so, it needs to both lift material out of the gravitational potential and expand it (effectively raising its enthalpy). These two contributions account for the observed decrease in luminosity\footnote{Note that this effect is not related to the pause itself but rather to the onset of the wind, when $L$ first exceeds $\Ledd$, and should therefore be a common feature across all PRE bursts. In fact, all lightcurves of pure He bursts in \citet{Yu.Weinberg2018} (see their Figure 12) also show a slowly descending flux throughout the wind, which can likely be attributed to the same effect.}. Second, near the end of the super-Eddington phase and before the decay, a bump in luminosity appears. This is related to the wind dying down, which ``returns" the gravitational energy and enthalpy that was required to sustain it back to the radiation. However, as discussed in the previous section, this part of the lightcurve is uncertain because of the wind becoming optically thin.

From the observational perspective, it could in principle be possible to infer the shape of the hydrogen profile from the lightcurve only. Assuming that the energy used to eject mass ($GMM_{\rm w}/R$ where $M_{\rm w}=4\pi R^2\yw$) is equal to a fraction $\eta$ of the observed burst energy (integrated luminosity which can be determined from the fluence if the distance to the source is known), one could infer $\yw$ from the lightcurve only. Then, given the total duration of the PRE, one could find the average $\Mdot$, and finally use Equation~\ref{eq:tpause} to obtain $\yc$. In our simulations, we find $\eta\approx0.31-0.37$, but this is likely to change for different ignition depths. We plan to study the energy budget of PRE bursts in more detail in future work. 

\section{Impact of changing the treatment of convection}\label{sec:otherconvection}

\begin{figure*}[ht!]
    \centering
    \includegraphics{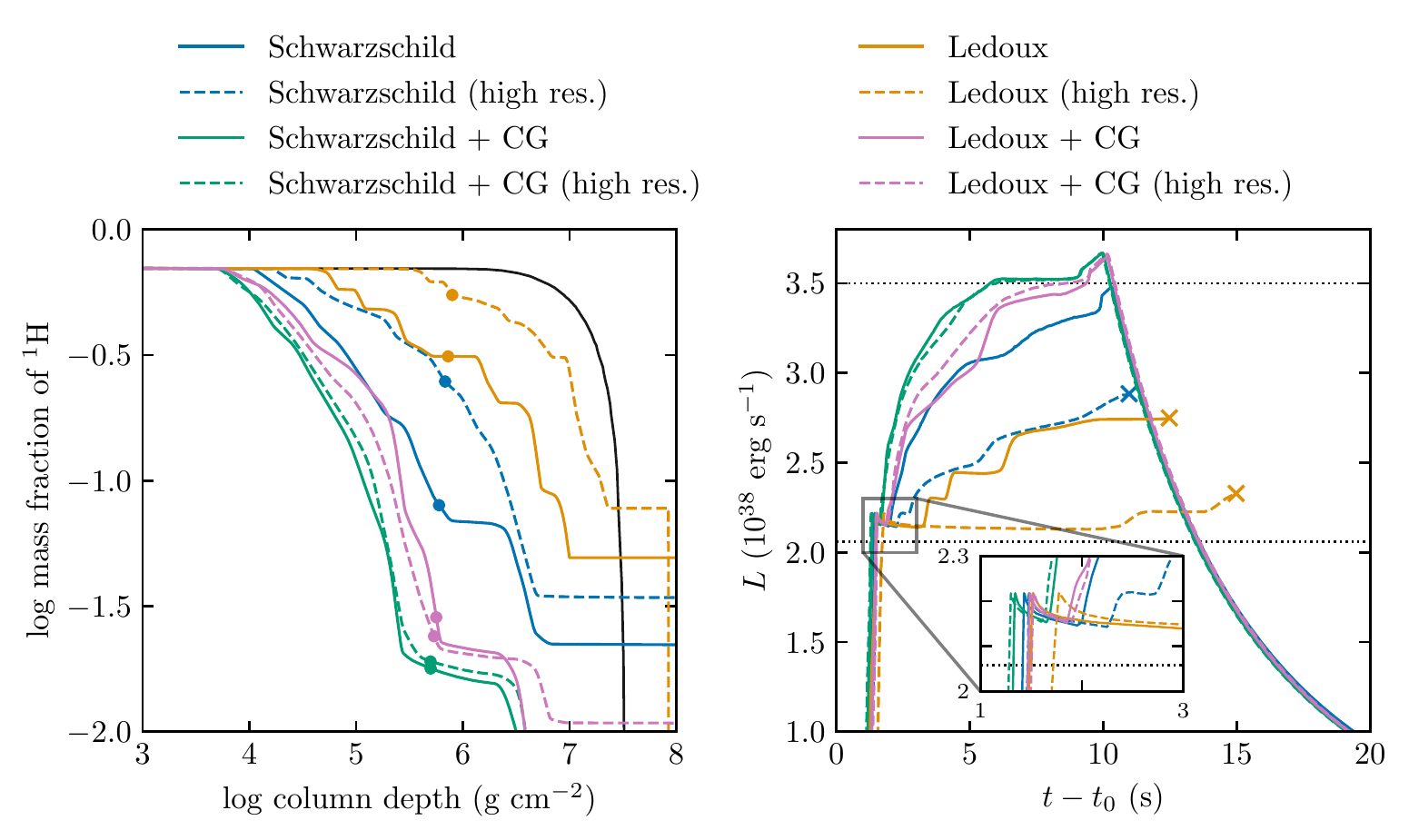}
    \caption{Results for all convective prescriptions and spatial resolutions tested in this work. \textit{Left}: Mass fraction of $^1$H in the atmosphere after the convective rise and before the ejection by winds. The circles label the locations of $\yw$ resulting from each simulations. The black line shows the hydrogen gradient at ignition, from which all runs start. \textit{Right}: Lightcurve of the bursts. The dotted lines mark the locations of $\Ledd$ for $X=0.7$ and $X=0$, as in Figure~\ref{fig:h1_lc}. Some simulations could not integrate through the decay phase and were stopped short at the ``x" symbols. The inset zooms in on the pauses.}
    \label{fig:all_models}
\end{figure*}

It is unlikely that the implementation of convection in Section \ref{sec:mesa} with the Schwarzschild criterion is an accurate representation of the true hydrodynamic phenomena. For one, the rapid nuclear burning induces local changes in composition towards heavier species, and locally increases the mean molecular weight $\mu$. The creation of $\mu$-gradients can either have a stabilizing or de-stabilizing effect on the thermal profile. This is especially relevant starting at the collision, where the growth of the He-rich convection zone will be inhibited by the H-rich material on top, as pointed out by \citet{Weinberg.Bildsten.ea2006}. They showed that a jump in temperature between the convection zone and the overlying radiative layer would develop in order to overcome the stabilization of the boundary due to composition. However, the effectiveness of the composition jump may be decreased by entrainment of fluid at the convective-radiative boundary which would erode the stabilizing composition gradient (e.g.~\citealt{Anders.Jermyn.ea2022_a}). Second, we may expect that the tiny radiative gaps that appear in between convection zones (see Section~\ref{subsec:ignition}) will be destroyed by some form of overshoot mixing. This is because, at the Schwarzschild convective boundary, the fluid parcel has zero acceleration by definition, but it may have enough inertia to continue rising and mix the fluid all the way to the next convective zone. 

To better understand the importance of these effects, we ran additional simulations, starting from the same model at ignition (Figure~\ref{fig:composition}, left panel), but changing the prescription for convection during the thermonuclear flash. We first ran a simulation using the Ledoux criterion instead of Schwarzschild for locating convective boundaries, which takes into account compositional gradients. When this is used, semiconvective and thermohaline mixing become available. For these, we set the dimensionless parameters $\alpha_{\rm sc}=0.1$, and $\alpha_{\rm th}=2$ \citep[\mesa{} uses these to determine diffusion coefficients, see][]{Paxton.Cantiello.ea2013}, following the \texttt{ns\_he} test suite problem in \mesa{}. Then, we ran simulations using both the Schwarzschild and Ledoux criteria where we also forced all radiative gaps of radial extent less than 10\% of the scale height to close and become convective instead\footnote{To implement this, we used \mesa{}'s existing optional routine (activated with the \texttt{min\_convective\_gap} control), but added an option to set a minimum pressure for the routine to operate, which we set to the pressure of the top of the iron substrate in order to prevent ``accidental" undershoot mixing with the iron. These additions are provided with the package linked in footnote \ref{footnote:zenodo}.}. This is meant to, in a simplified way, mimic the overshoot at the top of each convective zone. Finally, we tested the convergence of our simulations by running a high resolution version of every prescription. For these, the number of grid points during the flash phase was increased from ${\sim}5000$ to ${\sim}15000$ (the exact number varies in time and is adjusted by \mesa{} using the \texttt{mesh\_delta\_coeff} control).

We show in Figure~\ref{fig:all_models} the hydrogen composition profiles and lightcurves for all the simulations mentioned above. In each case, the main two features of our simplified model hold: \textit{1)} a larger $\yc$ results in a longer pause, and \textit{2)} a steeper hydrogen gradient leads to a faster rise in luminosity. The peak luminosity of the burst is then inversely proportional to $X(\yw)$, where $\yw\lesssim 10^6$ g cm$^{-2}$ is roughly constant across different bursts, and reaches the helium $\Ledd$ in half of the simulations. Moreover, lightcurves with higher peak luminosities have shorter PRE times -- this is because the fluence is conserved for a given ignition depth, no matter the exact the shape of the lightcurve. The slow decrease of the luminosity during the pause is also observed in every case. For runs which managed to integrate through the decay phase (ones which do not end in an ``x" in Figure~\ref{fig:all_models}), the bump at the end of the super-Eddington phase is also present. Once $\yw$ has been completely ejected, all models join on the same exponential cooling track.

While Figure~\ref{fig:all_models} demonstrates agreement with the basic model described in Section~\ref{sec:description}, it also clearly shows that the results, and in particular the observable lightcurve, depend on the prescription for convection. We cannot confidently claim that one prescription is more realistic than another, given the complex interactions between nuclear burning and mixing in this inherently multi-dimensional process, and so cannot predict lightcurves using these simulations. Nevertheless, we can assess the impact that different convective prescriptions have on the overall simulation. First, using the Ledoux criterion instead of the Schwarzschild criterion means that formerly convective regions become semi-convective instead, as the composition gradients stabilize the thermal profile. This effect inhibits the mixing, and, therefore (Section~\ref{subsec:ignition}), reduces the amount of hydrogen burned away. This can be seen in the left panel of Figure~\ref{fig:all_models} by comparing the orange and blue lines. Closing the radiative gaps between convective zones (``CG" in the figure legend) naturally has the opposite effect, with the mixing becoming stronger. The impact of spatial resolution is also interesting (compare solid and dashed lines in Figure~\ref{fig:all_models}). In the Schwarzschild and Ledoux runs, we obtain an increase in the total number of convective zones when increasing the resolution. Indeed, the addition of grid points in the convection zone allows it to split even more. Therefore, these runs are clearly not converged. However, in the CG runs, this splitting is effectively cancelled, as tiny zones are merged together at the end of every step, and we find good agreement between the models with different resolutions.

Another aspect to consider is the criterion for locating convective boundaries. In the presence of composition discontinuities, as is the case in our simulations, simplified applications of the Schwarzschild or Ledoux criteria can lead to unphysical scenarios and ultimately impede the growth of convective zones (see \citealt{Gabriel.Noels.ea2014} for an extensive discussion). In \citet{Paxton.Schwab.ea2018}, an optional ``predictive mixing'' scheme was implemented in {\mesa} to address this issue. This algorithm iteratively tests if the convective boundary is still correctly located once the material becomes fully mixed on the convective side, applying corrections otherwise. This may not be accurate in our situation where mixing also leads to rapid burning\footnote{For this same reason, the ``convective premixing'' scheme implemented in \citet{Paxton.Smolec.ea2019}, which works by iteratively instantaneously mixing and changing abundances in the cells surrounding the boundary, could not be used here.}. Nonetheless, we ran some exploratory simulations of the flash including predictive mixing. The results were very similar in all cases, except when using the Ledoux criterion without closing gaps. Then, the hydrogen abundances were reduced (mixing was enhanced) compared to the original model without predictive mixing, but not all the way to the models with the Schwarzschild criterion.

Finally, as mentioned in section \ref{subsec:ignition}, we found some large convective velocities resulting from the rapid nuclear burning, implying a departure from a hydrostatic envelope. We also find similar velocities in the other simulations presented in this section. However, the usual implementation of mixing-length theory does not factor in the time required to accelerate to these velocities. To explore this, we evaluated the timescale for convective acceleration at any given step $n$, given by \citet{Wood1974} as $\tau_{\rm a}\sim \ell/(v_{\rm c}^n+v_{\rm c}^{n+1})$, where $\ell$ is the mixing length and $v_{\rm c}$ is the convective velocity. If the actual timestep taken by the code $\Delta t=t^{n+1}-t^n$ is shorter than $\tau_a$, then velocities have increased too quickly. In our simulations, we found a few instances where $\Delta t/\tau_a \sim 10^{-3}$, but only during the precursor phase. This casts doubt on the existence and behaviour of this precursor. Alternatively, \mesa{} has the option to enable acceleration-limited convection \citep{Paxton.Marchant.ea2015}. The most recent version also includes a more complete theory of time-dependent convection \citep{Jermyn.Bauer.ea2022} which will also limit accelerations. However, our attempts at running simulations with either option were unfruitful, with the evolution eventually being driven to prohibitively short timesteps in trying to converge. Getting these models to converge and reach Eddington is a potential avenue for future work, as they are likely the most accurate way to model convection with rapid burning while remaining in one dimension.

\section{Summary and Discussion}\label{sec:discussion}

We have shown that variations in chemical composition in the envelope of neutron stars accreting mixed H/He fuel are reflected in the lightcurves of their PRE bursts. After the ignition of the thermonuclear runaway in the He-rich layer, a convective zone expands outwards and mixes the fuel with the overlying H-rich shell (Figure~\ref{fig:composition}). The resulting H abundance profile determines the shape of the lightcurve, namely the duration of initial pause and the subsequent slope in luminosity (Figure~\ref{fig:h1_lc}).
Due to convection, the mass of the layer with solar composition, $\yc\sim10^4-10^5$ g cm$^{-2}$, is much reduced compared to the initial $\yd\sim10^7$ g cm$^{-2}$ set by stable hydrogen burning during accretion (Equation \ref{eq:yd}). This results in a rapid ejection of a hydrogen-rich shell and a short observed pause on the order of 1 s or less (Equation \ref{eq:tpause}). Subsequently, the luminosity rises toward the helium Eddington luminosity as hydrogen depleted layers are exposed by the wind.

We find that the hydrogen profile in the envelope is sensitive to the details of convection and mixing following the collision with the H layer (Figures \ref{fig:kipp} and \ref{fig:kippH}). As a result, the exact shape of the lightcurve of a given event is uncertain as it depends on the choice of convective prescription and spatial resolution (Figure~\ref{fig:all_models}). The critical factor in setting the hydrogen profile is the efficiency of the mixing within the convective regions. However, this mixing is inhibited when the convection splits into many zones interspersed with radiative gaps. This splitting occurs even when ignoring compositional gradients (Schwarzschild criterion), suggesting that the culprit is the local energy deposition from rapid nuclear burning. Moreover, we found that increasing the spatial resolution of the simulations led to an increase in the number of zones and gaps, significantly reducing the efficiency of mixing, such that our simulations are not converged. This non-convergence however is mitigated by overshoot mixing at the top of convective zones, which we modeled in a simplified way by closing radiative gaps less than 10\% of the scale height.

Even disregarding problems related to splitting of the convection, a more fundamental issue stems from the approximate treatment of convection with mixing-length theory. In the collision event, nuclear burning releases energy on tens of microsecond timescales, which is close to or even shorter than local convective turnover times. This is in violation of the standard assumptions of mixing length theory. It also amplifies the differences between the Schwarzschild and Ledoux criterion, in contrast to situations with long dynamical timescales such as the main-sequence, where both criteria should lead to similar outcomes \citep{Anders.Jermyn.ea2022_a}, although we did find that an improved implementation of convective boundaries using predictive-mixing brought Ledoux closer to Schwarzschild. This timescale problem has been noted before in the context of late-stage evolution of massive population III stars. There, a helium burning convective region encroaches upon a hydrogen shell, mixing in protons which burn on timescales of hours to days, which is short compared to the month-long convective turnover times \citep{Marigo.Girardi.ea2001}. The proper modeling of these situations, which are also known as level-3 mixing or convective-reactive phases \citep{Herwig.Pignatari.ea2011}, continues to be an active area of research \citep[e.g.][]{Davis.Jones.ea2019,Clarkson.Herwig2021}, with a particular focus on multidimensional hydrodynamics simulations \citep{Woodward.Herwig.ea2014,Stephens.Herwig.ea2021}. 

Although the general shape of the lightcurves in our simulations agrees with the burst from \src{} reported by \citet{Bult.Jaisawal.ea2019}, there are some differences. In the observed burst, the pause is ${\sim}0.7$~s long, similar to our principal Schwarzschild run, suggesting a similar extent of the convection. However, the subsequent rise is rapid, reaching the helium Eddington luminosity in just ${\sim}1.3$~s. This would imply a mixing event which is strong enough to produce a very steep hydrogen gradient. Since the total pause plus rise duration is ${\sim}3$ times that of the pause, the hydrogen profile would have to go from\footnote{\citet{Goodwin.Galloway.ea2019} inferred a hydrogen mass-fraction $X_0\approx0.57^{+0.13}_{-0.14}$ for the companion, based on an analysis of Type I X-ray burst recurrence times and energetics. The observed ratio of peak to pause luminosities in \citet{Bult.Jaisawal.ea2019} favors the upper end of this range.} $X\approx 0.7$ to $X=0$ in the span of $\yc\approx10^4$~g cm$^{-2}$ to ${\approx}3\,\yc$. Or, if for example the mass-loss rate increases by a factor of 3 from the pause to the rise, as it does in our simulations (Section~\ref{subsec:wind}), then in the span of $\yc$ to ${\approx}9\,\yc$; in any case, the whole hydrogen gradient spans a decade in column depth at most. None of our simulations achieve this -- our fastest rise time is ${\gtrsim}6$~s for the Schwarzschild+CG model (in fact, factoring redshift, these times should be ${\sim}20-30$\% longer). Moreover, our Figure~\ref{fig:all_models} shows a general trend that steep hydrogen gradients are also associated with smaller $\yc$ values; to reproduce the rise seen in \src{}, we may need such strong mixing that it would push $\yc$ to very small columns and dissolve the pause entirely. 

One way to match the rapid rise but keep the pause duration the same as observed in \src{} could be to burn hydrogen more effectively with the same mixing efficiency and convective extent. In fact, our simulations do not model hydrogen burning completely, because we were limited to a small nuclear network which reached its end at $\iso{24}{Mg}$ prior to the wind launch (see right panel of Figure~\ref{fig:composition}). We investigated the effect of a larger network by running a simulation equivalent to our Schwarzschild run, but using \mesa{}'s \texttt{rp\_153.net} nuclear network, which includes isotopes up to $\iso{56}{Ni}$, until the wind launch. We found that the outer hydrogen profile was unchanged, with $\yc$ and initial hydrogen gradient staying the same. The effects of the additional hydrogen burning were limited to large columns $\gtrsim0.3\yw$. At these depths, hydrogen completely burned away, whereas with the smaller network a small amount ($\lesssim 0.1$ for the well-mixed models) remains. The lightcurve for such a burst would initially look the same as in our original Schwarzschild run (Figure~\ref{fig:h1_lc} bottom panel), but would continue rising all the way to the helium Eddington luminosity, instead of levelling of to ${\sim}90$\% of it. This suggests that additional burning is not the explanation for the rapid rise after the pause.

More observations of PRE bursts in the pure helium ignition regime would help to further understand and constrain the hydrogen ejection model. Note that previous PRE bursts from \src{}, observed with the Rossi X-ray Timing Explorer, have shown an increase in luminosity during the Eddington phase, but only by ${\sim}20$\% and on ${\sim}7$ s timescales \citep[see Figure 3 in ][]{Galloway.Goodwin.ea2017}. In these bursts, pauses are not clearly seen, which would indicate small values of $\yc$, although this could also be due to the choice of time bins used for the analysis. Such variations in the shape of the lightcurve (slow or fast luminosity increases) across different bursts from a single source may also imply that the dynamics of convection are very sensitive to initial conditions at ignition. Furthermore, a puzzling aspect of the burst reported in \citet{Bult.Jaisawal.ea2019} is the secondary peak following the PRE phase. This is unexplained by our hydrogen ejection model, and could instead require multidimensional effects.

On the theoretical side, the obvious next step in order to refine predictions for these bursts will be to improve the treatment of convection during the thermonuclear flash, in particular for the collision between the He and H layers. Due to the timescales involved and the limitations of mixing length theory, we know that only multidimensional hydrodynamical simulations can yield accurate results. This may pose a significant numerical challenge, although recent works by \citet{Malone.Zingale.ea2014} and \citet{Zingale.Malone.ea2015} have shown promising results in this direction, demonstrating the use of low-mach number hydrodynamics to model two and three-dimensional convection in thermonuclear explosions.

Finally, other improvements need also be made in the hydrodynamical part of the simulation in order to correctly model the super-Eddington wind. First, we faced numerical problems with ``staircases" in mass fractions leading to density inversions in the wind, which we simply smoothed out in this work. It would be interesting to investigate such density inversions as they propagate outward in future work. We also had issues at the end of the super-Eddington phase and collapse of the atmosphere. To properly model this part of the PRE, we will likely need hydrodynamical simulations which can handle optically thin radiative transfer as well as shocks (if our findings that infall velocities can be supersonic are correct). Hydrodynamical simulations would also be useful to model the super-Eddington winds in multiple dimensions, where the effects of rotation and magnetic fields could be taken into account. Lastly, for accurate observational predictions, it would be pertinent to include general relativistic corrections to the hydrodynamic equations, as they are known to result in larger photospheric radii \citep{Paczynski.Proszynski1986,Guichandut.Cumming.ea2021}.\\

We thank Alexander Heger for helpful discussions on convective-reactive mixing. We also thank Hang Yu for sharing results and analysis of his 2018 paper, and Rob Farmer for guidance on the \mesa{} source code. Finally, we thank the anonymous referee for helpful comments that improved this manuscript. S.G. is supported by an NSERC scholarship. This work was supported by NSERC Discovery Grant RGPIN-2017-04780. SG and AC are members of the Centre de Recherche en Astrophysique du Québec
(CRAQ). Simulations were ran on the
Graham cluster, operated
by the Digital Research Alliance of Canada.

\software{This work made use of the Python libraries \textit{NumPy} \citep{Harris.Millman.ea2020}, \textit{SciPy} \citep{Virtanen.Gommers.ea2020}, and \textit{Matplotlib} \citep{Hunter2007}. The \textit{py\_mesa\_reader} package \citep{Wolf2017_mesareader} was used for \mesa{} output files.}

\newpage
\bibliographystyle{aasjournal}
\bibliography{references} 
\end{document}